# MIDIS: Unveiling the Star Formation History in massive galaxies at $1 < z < 4.5$ with spectro-photometric analysis

M. Annunziatella[1] and P. G. P'erez-Gonz'alez[1] J. Álvarez-Márquez[1] L. Costantin[1] E. Iani[2] Á. Labiano[3]

P. Rinaldi[4] L. Boogaard[5] R. A. Meyer[5,6] G. "Ostlin[7] L. Colina[1] J. Melinder[7] I. Jermann[8,9]

S. Gillman[8,9] D. Langeroodi[10] J. Hjorth[10] A. Alonso-Herrero[11] A. Eckart[12] F. Walter[5] P. P. van der Werf[13]

A. Bik[7] F. Peißker[12] K. I. Caputi[2] M. García-Marín[14] G. Wright[15] T. R. Greve[8,9,16]

1. Centro de Astrobiología (CAB), CSIC-INTA, Ctra. de Ajalvir km 4, Torrejón de Ardoz, E-28850, Madrid, Spain
2. Kapteyn Astronomical Institute, University of Groningen, P.O. Box 800, 9700 AV Groningen, The Netherlands
3. European Space Agency (ESA), European Space Astronomy Centre (ESAC), Camino Bajo del Castillo s/n, 28692 Villanueva de la Cañada, Spain
4. Steward Observatory, University of Arizona, 933 North Cherry Avenue, Tucson, AZ 85721, USA
5. Max Planck Institut für Astronomie, Königstuhl 17, D-69117, Heidelberg, Germany
6. Department of Astronomy, University of Geneva, Chemin Pegasi 51, 1290 Versoix, Switzerland
7. Department of Astronomy, Oskar Klein Centre, Stockholm University, AlbaNova University Center, 10691 Stockholm, Sweden
8. DTU Space, Technical University of Denmark, Elektrovej 327, 2800 Kgs. Lyngby, Denmark
9. Cosmic Dawn Center (DAWN), Denmark
10. DARK, Niels Bohr Institute, University of Copenhagen, Jagtvej 155A, 2200 Copenhagen, Denmark
11. Centro de Astrobiología (CAB), CSIC-INTA, Camino Bajo del 1166 Castillo s/n, E-28692 Villanueva de la Cañada, Madrid, Spain
12. I.Physikalisches Institut der Universität zu Köln, Zülpicher Str. 77, 50937 Köln, Germany
13. Leiden Observatory, Leiden University, P.O. Box 9513, 2300 RA Leiden, The Netherlands
14. European Space Agency, Space Telescope Science Institute, Baltimore, Maryland, USA
15. UK Astronomy Technology Centre, Royal Observatory Edinburgh, Blackford Hill, Edinburgh EH9 3HJ, UK
16. Dept. of Physics and Astronomy, University College London, Gower Street, London WC1E 6BT, United Kingdom

August 23, 2025

**ABSTRACT**

*Context.* This paper investigates the star formation histories (SFHs) of a sample of massive galaxies ($M_\star \geq 10^{10}\ M_\odot$) in the redshift range $1 < z < 4.5$.
*Methods.* We analyze spectro-photometric data combining broadband photometry from HST and JWST with low-resolution grism spectroscopy from JWST/NIRISS, obtained as part of the MIDIS (MIRI Deep Imaging Survey) program. SFHs are derived through spectral energy distribution (SED) fitting using two independent codes, BAGPIPES and synthesizer, under various SFH assumptions. This approach enables a comprehensive assessment of the biases introduced by different modeling choices.
*Results.* The inclusion of NIRISS spectroscopy, even with its low resolution, significantly improves constraints on key physical parameters, such as the mass-weighted stellar age ($t_M$) and formation redshift ($z_{form}$), by narrowing their posterior distributions.
The massive galaxies in our sample exhibit rapid stellar mass assembly, forming 50% of their mass between $3 \leq z \leq 9$. The highest inferred formation redshifts are compatible with elevated star formation efficiencies ($\epsilon$) at early epochs.
Non-parametric SFHs generally imply an earlier and slower mass assembly compared to parametric forms, highlighting the sensitivity of inferred formation timescales to the chosen SFH model—particularly for galaxies at $z < 2$.
We find that quiescent galaxies are, on average, older ($t_M \sim 1.1$ Gyr) and assembled more rapidly at earlier times than their star-forming counterparts. These findings support the "downsizing" scenario, in which more massive and passive systems form earlier and more efficiently.

**Key words.** galaxies:general − galaxies: formation− galaxies: evolution − galaxies: high-redshift − galaxies: interactions

## 1. Introduction

Galaxies are intricate systems typically composed of multiple stellar populations. Observational constraints on their stellar mass assembly provide valuable insights into the physical processes that shape galaxy formation and evolution. Understanding the timescales on which these processes operate is crucial for addressing long-standing questions in galaxy evolution, such as when and how galaxies cease forming stars (i.e., quench; Schawinski et al. 2014; Schreiber et al. 2016; Carnall et al. 2018a), and what drives the observed bimodality in galaxy properties (e.g., Whitaker et al. 2012b; Muzzin et al. 2013).

In contrast to the hierarchical growth of dark matter halos predicted by ΛCDM models, observations suggest that the stellar components within these halos (i.e., galaxies) grow in a manner that is, at least to some extent, antihierarchical or top-down. In particular, the most massive galaxies observed in the local Universe appear to have assembled the bulk of their stellar mass rapidly and were already in place at early cosmic times (Pérez-González et al. 2008b; Marchesini et al. 2014; Forrest et al. 2020a).

The launch of JWST (Gardner 2023) has marked a major leap forward in infrared astronomy, enabling the detection of massive ($> 10^{10} M_\odot$) galaxies at increasingly earlier epochs (e.g., $z > 6$; Chworowsky et al. 2024; Shapley et al. 2025).

While targeting massive galaxies at high redshift is a powerful way to investigate their formation, this requires deeper observations and limits the sample size. A complimentary approach involves studying such galaxies at lower redshifts and reconstructing their star formation histories (SFHs) through spectral energy distribution (SED) modeling. This method has been shown to provide critical information on the timing of major star formation episodes and the processes that lead to their quenching (e.g., Carnall et al. 2018b; Iyer et al. 2019; Tacchella et al. 2022). However, SEDs are subject to significant and interrelated degeneracies among physical properties such as age, metallicity, and dust attenuation (e.g., Papovich et al. 2001; Lee et al. 2007; Conroy 2013). As a result, photometric observations alone are generally sensitive only to the most recent ∼1 Gyr of star formation, whereas the addition of spectroscopic data can extend sensitivity to earlier epochs (Chaves-Montero & Hearin 2020). Spectroscopic data provide key leverage in breaking these degeneracies. This is primarily because spectroscopy offers both emission line and continuum information, each probing different physical processes. Emission lines trace recent or ongoing star formation and nebular conditions (e.g., ionization state, gas metallicity), while the continuum shape and absorption features are sensitive to the underlying stellar populations, dust extinction, and stellar metallicity (e.g. Maraston 2005; Byler et al. 2017). The combination of these components allows for more accurate constraints on galaxy properties than photometry alone, especially when modeling complex star formation histories (e.g. Pacifici et al. 2012).

The advent of large spectroscopic surveys in the local Universe has greatly increased the availability of high signal-to-noise ratio (S/N) continuum spectra, allowing for tighter constraints on galaxy properties (e.g., Pacifici et al. 2012; Thomas et al. 2017). At higher redshifts, however, such high-quality data remained rare prior to JWST. Despite this limitation, several robust trends have been identified. At fixed redshift, lower-mass galaxies consistently exhibit younger stellar populations compared to more massive systems—a phenomenon known as "downsizing" (e.g., Gallazzi et al. 2005, 2014; Pacifici et al. 2016; Carnall et al. 2018a). In addition, at fixed stellar mass, the average formation redshift—defined as the redshift at which a galaxy has assembled 50% of its stellar mass—tends to decrease with decreasing observed redshift. This trend likely reflects the combined effects of several mechanisms, including the quenching of newly formed galaxies that join the red sequence (e.g., Brammer et al. 2011; Muzzin et al. 2013; Tomczak et al. 2014), galaxy mergers (e.g., Khochfar & Silk 2009; Khochfar et al. 2011; Emsellem et al. 2011), and episodes of rejuvenated star formation (e.g., Belli et al. 2017). The advent of JWST has finally allowed spectroscopic surveys targeting higher redshift massive galaxies (e.g. Carnall et al. 2024; Slob et al. 2024)

Despite their advantages, spectroscopic observations are significantly more resource-intensive than photometric ones, particularly when targeting faint, high-redshift galaxies. As a result, spectroscopic samples at early cosmic times have historically been limited in both size and scope. However, in this work, we demonstrate that even low-resolution spectroscopy from the JWST Near Infrared Imager and Slitless Spectrograph (NIRISS; Doyon et al. 2023) provides sufficient spectral coverage and sensitivity to significantly improve constraints on the star formation histories of massive galaxies. The com-bination of low-resolution (R=150) continuum and emission line information from JWST/NIRISS, together with broadband photometry, enables robust recovery of key SFH parameters, like mass-weighted ages and formation redshifts. We take advantage of these spectro-photometric data from JWST/NIRISS, in combination with available broadband and medium-band photometry, to investigate the evolution of massive galaxies in the redshift range $1 < z < 4.5$. To this end, we employ state-of-the-art modeling codes by exploring multiple SFHs.

Recent years have seen a significant effort to recover the 'true' SFH of galaxies. Traditionally, simple parametric form have been used to describe it. Among the most commonly used forms are the delayed exponentially declining (e.g. Carnall et al. 2019). However, such models often fail to capture the diversity and complexity of galaxy growth, especially at high redshift. To address this, more sophisticated parametric forms have been introduced. For instance, two-component models, combining an old population (modeled with an exponentially declining SFH) and a recent burst (often described with a double power law), have been successfully employed to model the SFHs of post-starburst galaxies (e.g., Carnall et al. 2018a). Recent years have seen an increase in the use of more flexible methods, such as non-parametric SFHs, which adopt a series of constant star-formation periods, divided into fixed or flexible time intervals (e.g. Iyer et al. 2019; Leja et al. 2019).

In this work, use two state-of-the-art codes: `BAGPIPES` (Carnall et al. 2018a, Bayesian Analysis of Galaxies for Physical Inference and Parameter Estimation) and `synthesizer` (Pérez-González et al. 2003a) with different assumptions on the SFH to describe the evolution of our massive galaxies. The paper is organized as follows: Section 2 describes the NIRISS observations and data reduction process; Section 3 outlines our sample selection; Section 3.1 details our spectro-photometric modeling approach; and Section 4 presents our main results.

Throughout the paper, we assume $\Omega_M = 0.3, \Omega_\lambda = 0.7, H_0 = 70 \mathrm{kms^{-1}Mpc^{-1}}$ and AB magnitudes (Gunn et al. 1987).

## 2. Data description

The analyses in this paper are based on NIRISS data taken in parallel with the MIRI Deep Imaging Survey (MIDIS; Östlin et al. 2025) in the Hubble ultra-deep field (HUDF) conducted by the MIRI European Consortium GTO program. Among its many results, MIDIS has played a key role in the discovery of previously undetected faint galaxies (e.g., Pérez-González et al. 2023, 2025), the first little red dot with clear detection of its host galaxy (e.g., Iani et al. 2025; Rinaldi et al. 2025b), and the detailed characterization of the role of the emitter at the epoch of reionization properties of distant galaxies (e.g., Rinaldi et al. 2023, 2024). In addition, MIDIS observations have provided new insights into galaxy morphologies at high redshift (e.g., Boogaard et al. 2024; Costantin et al. 2025).

The NIRISS observations cover a southern portion of the GOODS-S field and consist of 20 h (including overheads) of direct imaging and grism observations in the three bands F115W, F150W, and F200W. This resulted in 1112s,1112s and 2232s on-source direct images in F115W, F150W and F200W, and ∼ 5.3h grism spectra in each band. These data were reduced using the official JWST pipeline version 1.8.4 (pmap=1019). In addition to the default pipeline stages, which include snowball correction, we also applied a background homogenization algorithm prior to obtaining the final mosaics, including 1/f- noise removal as in Bagley et al. (2023).

The images were then registered to the same reference frame of the World Coordinate System using the Hubble Legacy Field (HLF) catalog in Whitaker et al. (2019), based on Gaia DR2 (Gaia Collaboration et al. 2016, 2018). In all three images, we then find a median offset of $\Delta\alpha = -0.001\ arcsec$ and $\Delta\delta = -0.002\ arcsec$. We left the pixel scale to the nominal NIRISS pixel scale (0.065 $''/pixel$). The $3\sigma$ depth of the direct images observations (measured in circular apertures of 0.2$''$ diameter) is ~ 28.6, 28.7, 29.1 AB, in F115W, F150W and F200W, respectively. The $3\sigma$ line flux sensitivity of the grism is ~ 1.5e − 18 erg s$^{-1}$cm$^{-2}$. The MIDIS parallel observations target a region (hereafter MIDIS-P3) of the GOODS-S field covered by HST ACS and WFC3 from the Hubble Legacy Field (HLF). In this paper, we use, in addition to our NIRISS data, the HLF v2 images for all filters from Whitaker et al. (2019). A small portion (~ 1/6th) of P3 overlaps with the NIRISS parallel observations of the JWST Extragalactic Medium-band Survey (GO) program JWST Extragalactic (JEMS, PID: 1963; Williams et al. 2023) in the two medium bands F430M and F480M. The NIRISS images from JEMS were reduced in the same way as for the MIDIS images.

Figure 1 shows the position of MIDIS-P3 in the GOODS-S field with respect to other JWST programs. It also shows a RGB color image of P3 obtained using the three direct images in F115W, F150W, and F200W.

Spectra extraction was performed using the official JWST pipeline. We added an extra step between stage 1 and stage 2 for background removal and homogenization on the 2D grism files. The 2D spectra in stage 2 were extracted in a region that corresponds to the Kron apertures (Kron 1980) identified in Section 3.

## 3. Sample selection

For this work, we selected galaxies previously detected in the CANDELS (Grogin et al. 2011; Koekemoer et al. 2011) with redshift $z \geq 1$, and stellar masses from Santini et al. 2015 log(M/M$_\odot$) $\geq$ 10.0. Figure 3 shows the mass and redshift distribution of our targets. Some (~ 10%) of the selected galaxies have secure spectroscopic redshift from different works (e.g. Bacon et al. 2023; Santini et al. 2015, and references therein). As P3 is part of the larger CDFS field, we also checked whether for these galaxies there were any spectra available from the VAN-DELS collaboration (Garilli et al. 2021), however, this is not the case. Our final sample consists of 47 galaxies.

After we identified our galaxy sample, we measured their photometry in the NIRISS images and re-measured their fluxes in the HST ones from HLF by fixing centers, shapes, and sizes to those of the Kron apertures derived for NIRISS. Finally, we cross-match this catalog to the multi-wavelength catalog from CANDELS in the GOODS-S field (from the *Rainbow* database[1] Pérez-González et al. 2008a; Barro et al. 2011) to obtain photometry in the IRAC and MIPS filter (Barro et al. 2011). The final catalog contains photometry in more than 20 filters.

Figure 2 shows the RGB images and NIRISS JHK spectra for a selection of sources in our sample.

From both the photometry and the final 1D spectra in the three NIRISS filters, we derived the spectro-photometric redshifts using the eazy-py code (Brammer 2021). As shown in Figure 3, we have a limited number of objects with independent and secure spectroscopic redshift measurements in our sample

(12). With respect to these objects, we obtain a $\sigma_{NMAD} = 0.001$[2] and no outliers (i.e. objects with $\Delta z/(1 + z) > 0.15$). Of the objects in our sample without previous redshift measurements, five galaxies have strong, identifiable emission lines (e.g. top and middle galaxies in Figure 2). Notably, the most distant galaxy in our redshift sample ($z = 4.564 \pm 0.001$) has been identified through the [O 2]$\lambda3727$ line. Our redshifts agree very well with the photometric redshifts previously determined from the CANDELS photometry. With respect to previous photometric redshifts, CANDELS redshifts have an outlier fraction (i.e. the fraction of objects with $\Delta z/(1 + z) > 0.15$) of 15%.

The galaxies in our sample span a redshift range between $1.0 \leq z \leq 4.5$. Figure 4, shows the distribution of our galaxies in the rest-frame $U - V$ vs $V - J$ diagram, hereafter UVJ, as obtained by eazy-py. Most (81%) of our galaxies lies in the star-forming region of the diagram, 19% of them are quiescent.

### 3.1. Spectro-photometric Modeling

After we obtained the spectro-photometric redshifts from *eazy-py*, we derive galaxy properties through Spectral Energy distribution (SED) fitting of both photometry and spectroscopy. We decided to use two different codes: BAGPIPES (Carnall et al. 2018a, Bayesian Analysis of Galaxies for Physical Inference and Parameter Estimation) and synthesizer (Pérez-González et al. 2003a). BAGPIPES is a Stellar Population Synthesis (SPS) modeling package built on the updated BC03 Bruzual & Charlot (2003) spectral library with the 2016 version of the MILES library (Sánchez-Blázquez et al. 2006). It is built on a Kroupa IMF (Kroupa 2001) assumption and uses a Multi-Nest nested sampling algorithm (Feroz et al. 2019) to produce posterior distributions of physical parameters. In our run, we also take into account the resolution NIRISS grism. The attenuation law is modeled with Calzetti et al. (2000). For this code, we assumed three different Star Formation Histories (SFHs): a single stellar population described by a delayed-exponential function, a non-parametric SFH and two stellar populations. The two-population model is characterized by an exponential declining model for the old stellar population plus a double power-law that describes the more recent bursts of star formation. For the non-parametric SFH we used the continuity model from Leja et al. (2019) which fits directly for $\Delta\log(SFR)$ between adjacent time bins. This prior explicitly weights against sharp transitions in SFR(t). We assume bins logarithmically spaced from 0 to the age of the universe at the redshift we observed the object. In case of two stellar populations, the older component is characterized by a delayed exponential function, while the SFH of the younger component is a burst characterized as a double power law.

The synthesizer code (Pérez-González et al. 2003b, 2008a) assumes a Chabrier IMF (Chabrier 2003) with mass limits between 0.1 and 100 $M_\odot$. The attenuation law is modeled with Calzetti et al. (2000). For synthesizer we considered a single stellar population characterized by a delayed exponential SFH, and two stellar populations, characterized by two delayed exponentials.

For BAGPIPES, a summary of the free parameters and their allowed ranges in the SED-fitting procedure for all SED fitting codes and for the different choice of SFHs is reported in Table 1. For synthesizer, we adopted the same ranges whenever possible. Figure 5 shows the full spectrophotometric modeling of one of the galaxy in our sample.

---

[2] $\sigma_{NMAD} = 1.48 \times median(\frac{\delta z - median(\delta z)}{1 + z_{spec}})$

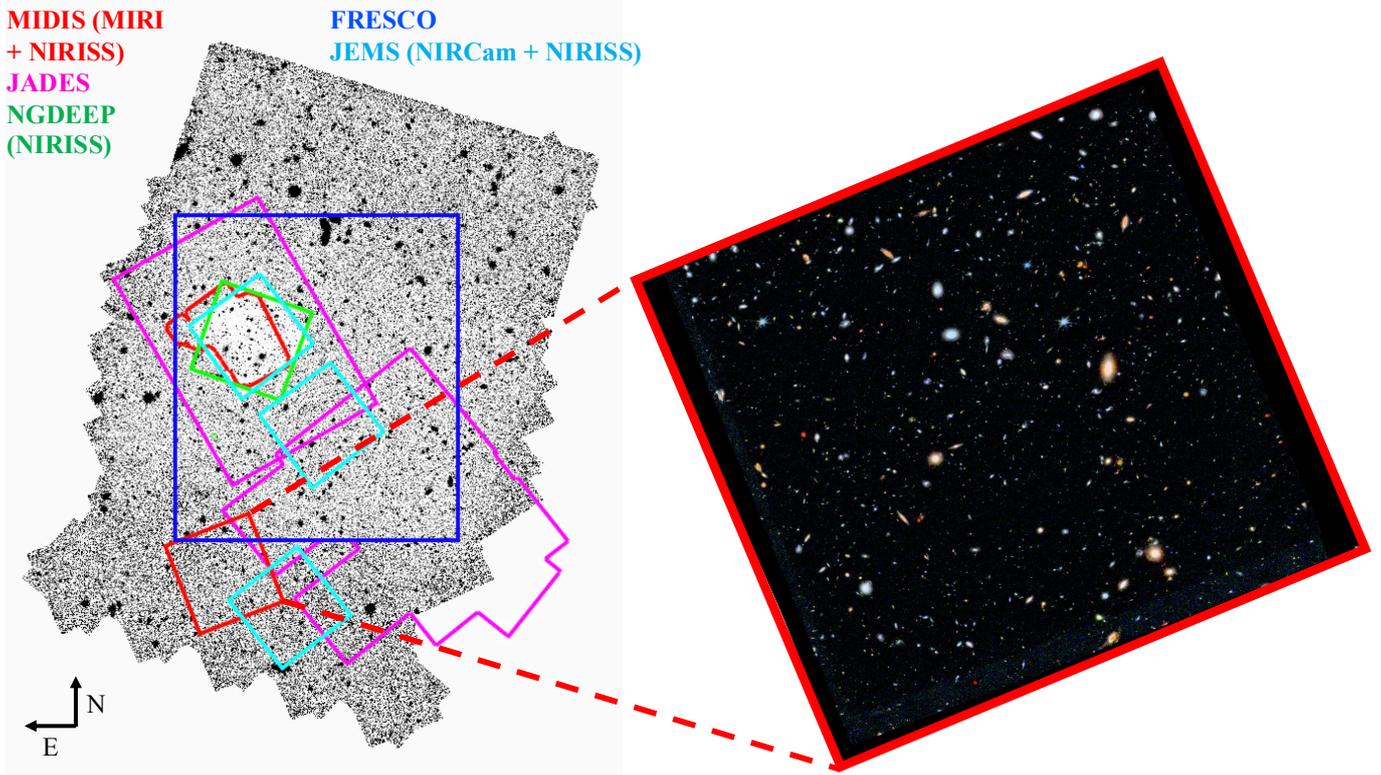

**Fig. 1.** Image of GOODS-S in $H_{160}$ from CANDELS. Over-plotted the regions observed by JWST by different programs. In red two out of three pointings of the MIDIS survey (P1 and P3). The red insert shows a color composite RGB image of the MIDIS-P3 field obtained using the three direct images in F115W, F150W, and F200W.

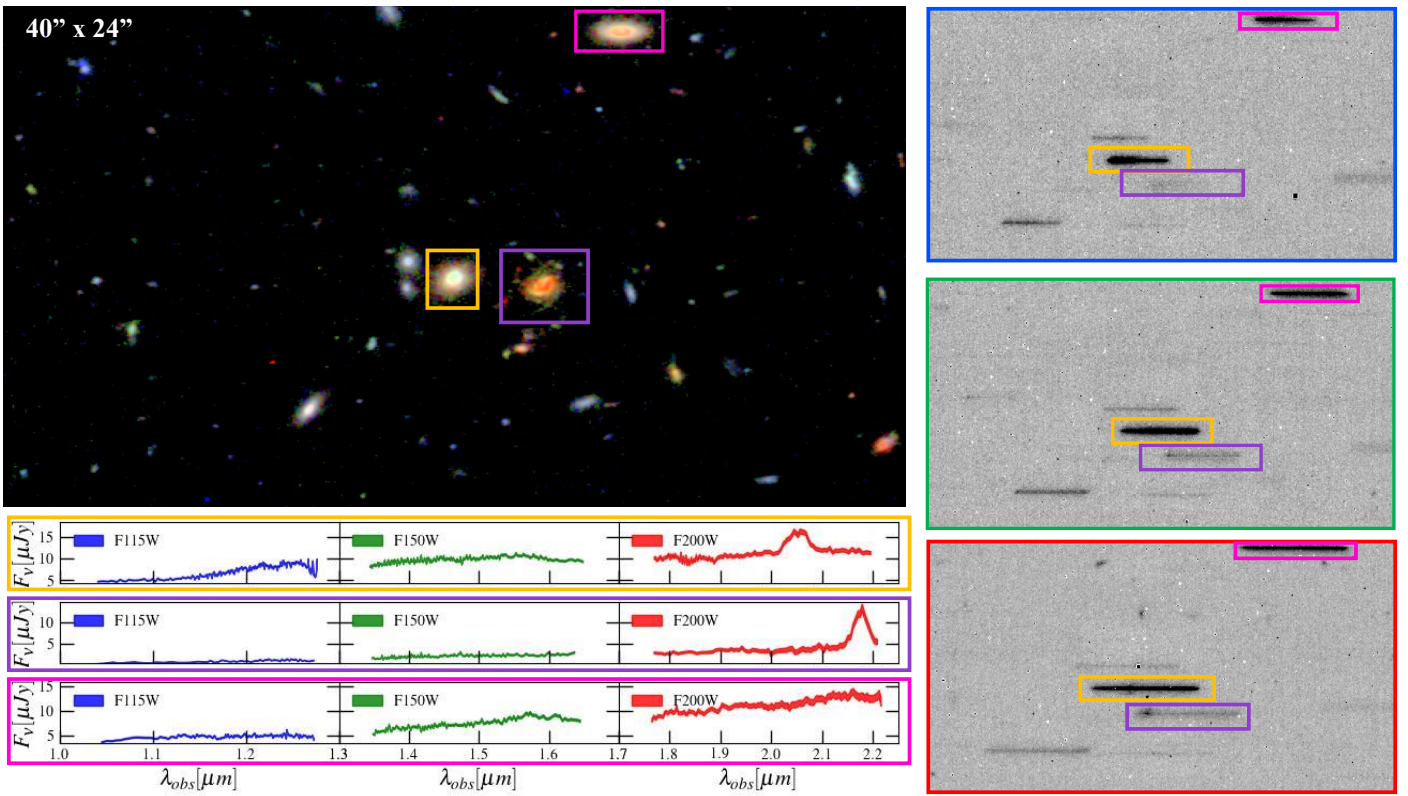

**Fig. 2.** On the top-left an RGB cutout of P3, centered on one of the galaxies in the sample. The three squares highlight three massive galaxies in the sample. On the right a single GRISM exposure containing the same objects, in the three filters and for one (GR150C) rotation. On the bottom, the extracted spectra with $3\sigma$ errors (shaded areas) of the three galaxies.

**Table 1.** Free parameters and their allowed ranges used in the SED-fitting procedure for all choices of SFH.

| Parameter | 1 POP | Non-parametric | 2 POP | Units |
|---|---|---|---|---|
| Time-scale ($\tau$) | (100, 1000 ) | | (100, 1000 ) | Myr |
| Dust attenuation ($A_V$) | (0, 8) | (0, 8) | (0, 8) | mag |
| Metallicity (Z) | (0.01, 2.5) | (0.01, 2.5) | (0.01, 2.5) | $Z_\odot$ |
| $\mathrm{Log_{10}}$ of the Nebular ionization parameter (logU) | (-4, -2) | (-4, -2) | (-4, -2) | |
| Number of age bins in log-space | | 10 | | |
| Burst strength | | | (0, 1) | |

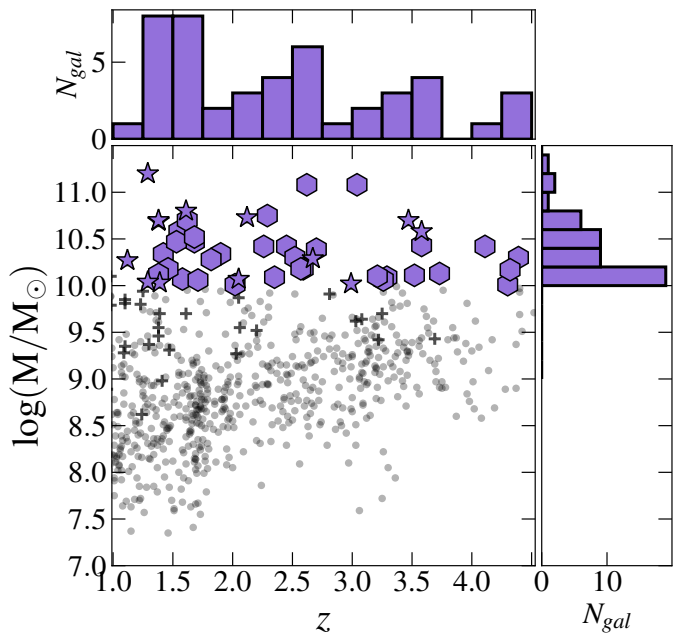

**Fig. 3.** Distribution in mass and redshift from the CANDELS catalog of all galaxies previously detected in P3. Black points are galaxies with photometric redshift and black crosses galaxies with spectroscopic redshift. In purple our selected galaxy sample, with hexagons representing galaxies with photometric redshifts and stars marking galaxies with spectroscopic ones.

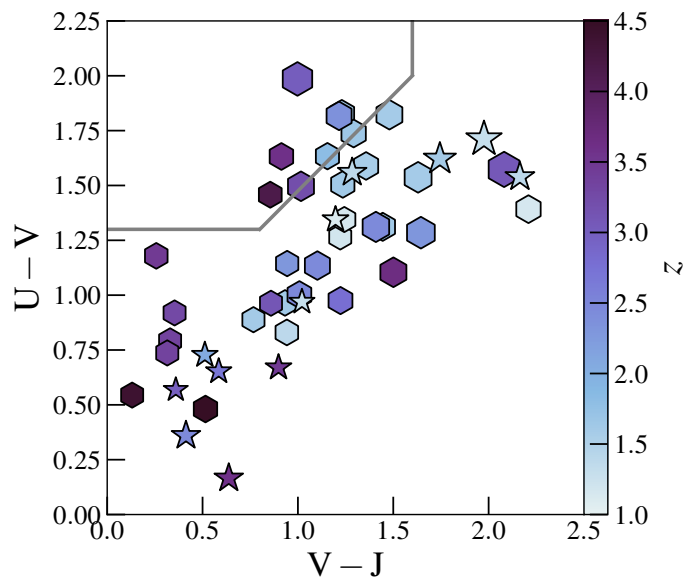

**Fig. 4.** UVJ diagram for the galaxies in our sample color-coded according to their redshift. The size of the points is proportional to the stellar mass. The symbols have the same meaning as in Figure 3. The lines mark the quiescent (top) and star-forming part of the diagram (Whitaker et al. 2012a).

In this work, we are interested in when massive galaxies start to assemble their stellar mass, how rapidly they do it, and how rapidly they quench, i.e. their star formation is halted. For this reason, we focus on a sub-sample of parameters among those that can be derived from the SED fitting. These parameters are: stellar mass, star formation rate, mass-weighted age, the redshift at which the galaxies form 50% of their stellar mass (i.e. formation redshift), and the time and redshifts at which galaxies form a fixed stellar mass ($10^8 - 10^9 M_\odot$).

We adopted the surviving mass as the stellar mass throughout this work. We define the mass-weighted age as follows:

$$\bar{t}_M = \frac{\int SFR(t) \times t \, dt}{\int SFR(t) dt},$$ (1)

where SFR(t) is the star formation rate as a function of time. We call formation redshift ($z_{form}$) the redshift corresponding to the mass-weighted age of the galaxy. In particular, first we translate the mass-weighted age to age of the Universe, and from this we identify the formation redshift. This parameter corresponds to the redshift at which a galaxy has assembled 50% of its stellar mass.

To understand how rapidly these galaxies assemble their stel-lar mass, we use, as an indicator of formation timescale, the time and the redshift at which galaxies have assembled their first $10^8$ and $10^9 M_\odot$ respectively ($t(10^8 M_\odot)$, $t(10^9 M_\odot)$, $z(10^8 M_\odot)$, $z(10^9 M_\odot)$).

The parameters defined above depend on a priori assumptions on SFHs and possibly on the peculiarities of different codes. Table 2 shows the distributions ($50th_{16th}^{84th}$) of the properties discussed in the paper for different classes of galaxies and with all codes. The same convention ($50th_{16th}^{84th}$) is used throughout the paper.

In Appendix A, we show how all the parameters mentioned above compare for the two codes and different SFH assumptions mentioned above.

### 3.2. The fiducial model and the importance of NIRISS low-resolution spectroscopy

To assess which model best matches our data, we begin by visually examining the stacked SEDs of quiescent and star-forming galaxies (Figure 6). Individual SEDs are shifted to the rest-frame before stacking, and the resulting median stack is redshifted to the sample's average redshift. For quiescent galaxies, the available data do not allow us to clearly distinguish between the different models, as the overall SED shapes and continuum levels are similar across all runs. However, in the case of star-forming galaxies, the right panel of Figure 6 reveals that neither of the synthesizer runs adequately reproduces the H$\alpha$ emission lines observed in the NIRISS spectra of a subset of the star-forming sample. This is also evident for the star-forming galaxy in the bottom panels of Figure 5. These discrepancies suggest that the emission line modeling and/or star formation histories implemented in the synthesizer runs are insufficient to capture all spectroscopic features of these galaxies. Consequently, we selected our fiducial models from the BAGPIPES runs, which also allow more flexibility in modeling complex star formation histories.

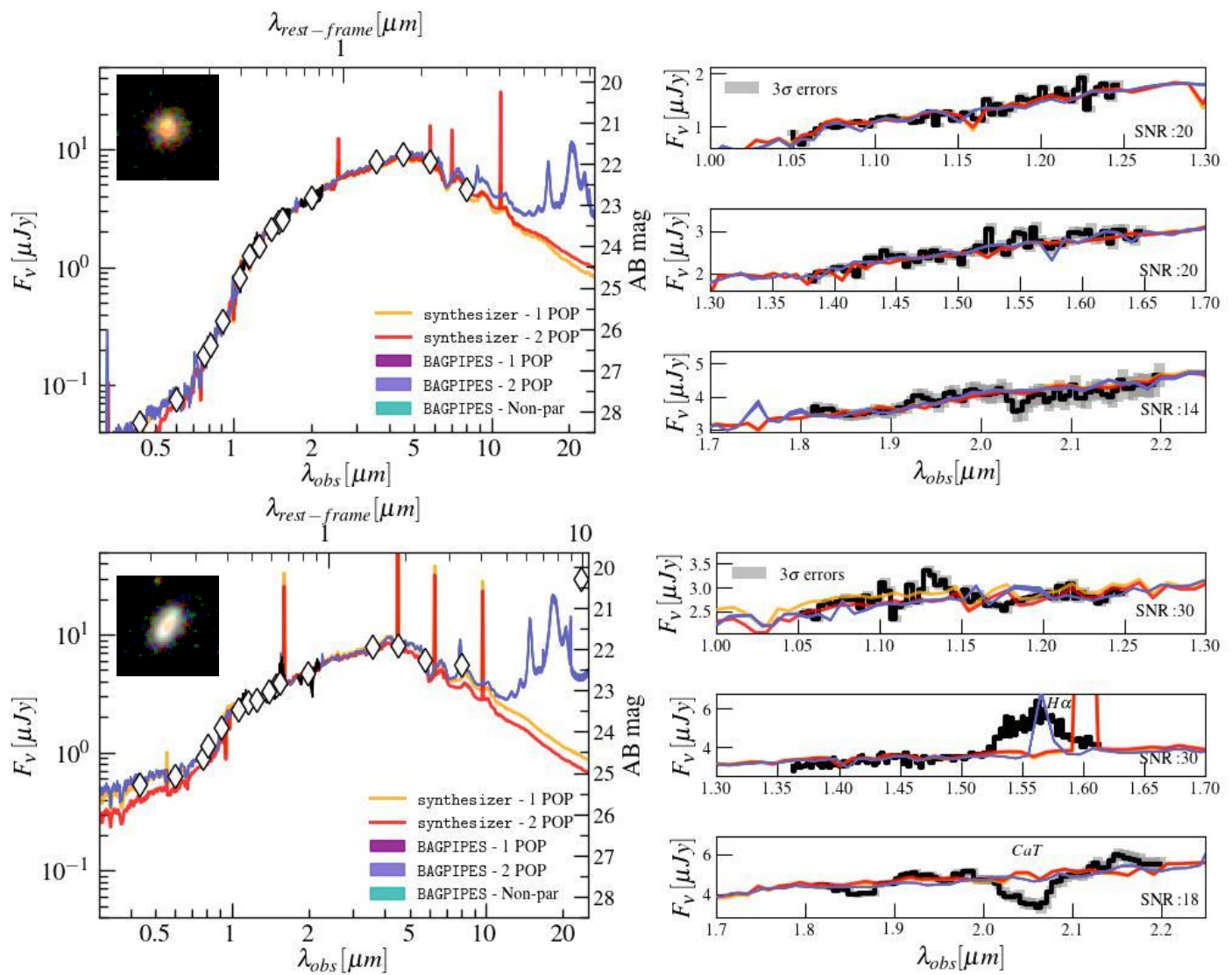

**Fig. 5.** Spectro-photometric SED fitting of two galaxies (a quiescent and a star-forming one) in our sample. The lines are the best-fit SEDs from different codes and SFH choice. Photometric data are shown as white diamonds. The error-bars are smaller than the size of the points. In black the 1D spectra in F115W, F150W and F200W. The right panels show a zoom-in of the spectra in linear scale. The $3\sigma$ errors are plotted as shaded Grey areas. We report the median S/N ratio of the 1D spectra.

We computed the reduced $\chi^2$ values consistently across all model configurations by including both photometric and spectroscopic data in the fits. Among the various runs, the configuration that yields the lowest median reduced $\chi^2$ is the BAGPIPES run with two stellar populations. To quantify the relative performance of each model, we computed the median $\chi^2$ ratio between each alternative run and the BAGPIPES 2-POP model. The resulting median ratios are 1.04, 2.28, 1.28, and 1.29 for the BAGPIPES 1-POP, BAGPIPES with a non-parametric star formation history, synthesizer 1-POP, and synthesizer 2-POP runs, respectively. Based on these results, we adopt the **BAGPIPES –**

**2POP** configuration as our fiducial model throughout the remainder of this analysis. For completeness, we also report the results obtained from the BAGPIPES – 1POP and BAGPIPES – Non-parametric configurations in the following sections, enabling comparisons and illustrating the impact of model assumptions on the inferred galaxy properties.

To evaluate the specific contribution of the NIRISS spectroscopy to constraining the physical properties of individual galaxies, we performed an additional BAGPIPES – 2POP run using only the photometric data. By comparing these results with those from our fiducial run, we aim to quantify the added value of the NIRISS spectra. Given our primary interest in understanding the assembly history of stellar mass in galaxies, we focus on the distributions of $t_M$ and $z_{form}$ as a function of total formed stellar mass. These distributions, derived from the Bayesian posterior samples provided by BAGPIPES, are shown for two representative galaxies in Figure 7, which correspond to those previously presented in Figure 5. Across the entire sample, we find that including the NIRISS spectra results in significantly tighter constraints on the derived parameters, as evidenced by narrower pos-terior distributions in the $t_M$–mass and $z_{form}$–mass planes. This tightening of the parameter space reflects the additional information content provided by the spectral data, particularly in terms of emission and absorption line features that are sensitive to recent and intermediate-age star formation. The improvement is clearly visible for the two example galaxies shown in Figure 7, where the inclusion of spectra leads to a marked reduction in the uncertainty of both stellar mass and formation timescales. The bottom panels of Figure 7 show that, the best-fit values for stellar mass, $t_M$, and $z_{form}$ shift slightly when spectroscopy is included, although they remain consistent within $1\sigma$ with the values derived from the photometry-only runs. This indicates that the spectroscopic data not only improve precision but also validate the overall robustness of the photometric estimates. As a result, we conclude that the addition of NIRISS spectroscopy, even with its low-resolution, enhances our ability to constrain the star formation histories of high-redshift galaxies.

**Table 2.** Average SED fitting properties of the galaxies in our sample for different codes. The fiducial model is highlighted in purple.

| Sample | Parameter | Codes | | | | |
|---|---|---|---|---|---|---|
| | | BP-1 POP | BP-2 POP | BP-Non-parametric | sy-1 POP | sy-2 POP |
| All galaxies ($\bar{z} = 2.4$) | $\log(M/M_\odot)$ | $10.5^{10.7}_{10.1}$ | $10.5^{10.9}_{10.1}$ | $10.5^{10.8}_{10.0}$ | $10.4^{10.7}_{10.1}$ | $10.4^{10.7}_{10.1}$ |
| | $\bar{t}_{M-w}$ [Gyr] | $0.4^{1.1}_{0.1}$ | $0.8^{2.1}_{0.3}$ | $0.6^{1.5}_{0.1}$ | $0.2^{1.0}_{0.0}$ | $0.7^{1.7}_{0.4}$ |
| | $z_{form}$ | $2.8^{4.8}_{1.7}$ | $3.9^{5.6}_{2.5}$ | $3.7^{4.3}_{2.5}$ | $2.8^{4.3}_{1.7}$ | $3.9^{5.4}_{2.4}$ |
| | $z_{M,*=10^8 M_\odot}$ | $3.8^{5.8}_{1.7}$ | $6.6^{8.9}_{1.9}$ | $9.5^{9.9}_{3.7}$ | $3.1^{5.5}_{1.9}$ | $6.0^{9.0}_{3.3}$ |
| | $z_{M,*=10^9 M_\odot}$ | $3.8^{5.7}_{1.8}$ | $5.9^{8.2}_{4.0}$ | $6.8^{9.5}_{2.8}$ | $3.1^{5.5}_{1.9}$ | $5.7^{8.1}_{3.2}$ |
| | SFR [$M_\odot yr^{-1}$] | $20^{247}_{0}$ | $62^{501}_{0}$ | $18^{242}_{1}$ | $29^{1591}_{5}$ | $236^{1748}_{12}$ |
| Lower-$z$ ($1 \leq z < 2$) ($\bar{z} = 1.6$) | $\log(M/M_\odot)$ | $10.4^{10.7}_{10.2}$ | $10.4^{10.8}_{10.3}$ | $10.5^{10.8}_{10.3}$ | $10.4^{10.6}_{10.1}$ | $10.4^{10.6}_{10.1}$ |
| | $\bar{t}_{M-w}$ [Gyr] | $0.9^{1.3}_{0.3}$ | $2.0^{3.0}_{0.3}$ | $2.2^{2.8}_{0.9}$ | $0.6^{1.7}_{0.1}$ | $1.4^{1.9}_{0.8}$ |
| | $z_{form}$ | $1.8^{2.7}_{1.6}$ | $3.4^{5.0}_{2.5}$ | $3.6^{3.9}_{2.9}$ | $1.7^{2.8}_{1.6}$ | $2.6^{3.8}_{1.8}$ |
| | $z_{M,*=10^8 M_\odot}$ | $1.9^{3.9}_{1.2}$ | $4.7^{8.7}_{1.8}$ | $9.8^{9.9}_{4.1}$ | $1.8^{3.1}_{1.6}$ | $3.5^{6.9}_{2.7}$ |
| | $z_{M,*=10^9 M_\odot}$ | $1.8^{3.8}_{1.7}$ | $4.5^{6.5}_{2.5}$ | $9.2^{9.4}_{4.0}$ | $1.8^{3.1}_{1.6}$ | $3.4^{6.7}_{2.6}$ |
| | SFR [$M_\odot yr^{-1}$] | $6^{11}_{1}$ | $5^{294}_{0}$ | $4^{29}_{0}$ | $3^{90}_{0}$ | $23^{183}_{0}$ |
| Higher-$z$ ($z \geq 2$) ($\bar{z} = 3.1$) | $\log(M/M_\odot)$ | $10.6^{10.7}_{10.1}$ | $10.6^{11.0}_{10.1}$ | $10.5^{10.8}_{10.2}$ | $10.3^{10.7}_{10.0}$ | $10.3^{10.7}_{9.6}$ |
| | $\bar{t}_{M-w}$ [Gyr] | $0.2^{0.9}_{0.1}$ | $0.5^{0.9}_{0.3}$ | $0.2^{1.0}_{0.1}$ | $0.1^{0.5}_{0.0}$ | $0.5^{1.1}_{0.4}$ |
| | $z_{form}$ | $3.8^{6.1}_{2.6}$ | $4.0^{5.6}_{3.9}$ | $3.7^{5.2}_{2.4}$ | $3.6^{4.8}_{2.6}$ | $4.6^{5.9}_{2.7}$ |
| | $z_{M,*=10^8 M_\odot}$ | $4.2^{6.1}_{2.9}$ | $7.3^{8.9}_{3.7}$ | $8.8^{9.3}_{4.2}$ | $3.7^{5.8}_{2.2}$ | $6.7^{9.9}_{4.6}$ |
| | $z_{M,*=10^9 M_\odot}$ | $4.2^{6.1}_{2.9}$ | $6.6^{8.4}_{4.3}$ | $4.4^{5.3}_{2.8}$ | $3.7^{5.7}_{2.2}$ | $6.2^{9.1}_{4.2}$ |
| | SFR [$M_\odot yr^{-1}$] | $30^{597}_{1}$ | $90^{573}_{2}$ | $34^{354}_{1}$ | $106^{2267}_{9}$ | $589^{2255}_{132}$ |
| Quiescent ($\bar{z} = 2.9$) | $\log(M/M_\odot)$ | $10.5^{10.6}_{10.3}$ | $10.5^{10.6}_{10.3}$ | $10.5^{10.7}_{10.4}$ | $10.4^{10.6}_{10.1}$ | $10.5^{10.6}_{10.3}$ |
| | $\bar{t}_{M-w}$ [Gyr] | $1.1^{1.7}_{1.0}$ | $1.1^{2.0}_{0.9}$ | $1.0^{2.2}_{0.6}$ | $0.8^{1.7}_{0.1}$ | $1.1^{1.6}_{0.9}$ |
| | $z_{form}$ | $4.8^{6.1}_{2.4}$ | $4.1^{7.4}_{3.5}$ | $4.2^{5.5}_{3.3}$ | $3.9^{5.6}_{3.2}$ | $4.2^{6.1}_{1.9}$ |
| | $z_{M,*=10^8 M_\odot}$ | $5.9^{7.4}_{2.9}$ | $6.8^{8.6}_{4.6}$ | $9.8^{9.9}_{3.4}$ | $4.2^{7.7}_{3.2}$ | $5.7^{10.8}_{3.2}$ |
| | $z_{M,*=10^9 M_\odot}$ | $5.8^{7.3}_{2.8}$ | $5.7^{8.0}_{4.4}$ | $9.2^{9.6}_{7.0}$ | $4.1^{7.4}_{2.7}$ | $5.4^{10.7}_{3.2}$ |
| | SFR [$M_\odot yr^{-1}$] | $1^{6}_{0}$ | $0^{1}_{0}$ | $0^{1}_{0}$ | $3^{41}_{0}$ | $17^{567}_{0}$ |
| Star-forming ($\bar{z} = 2.4$) | $\log(M/M_\odot)$ | $10.4^{10.7}_{10.1}$ | $10.5^{11.0}_{10.2}$ | $10.5^{10.9}_{10.2}$ | $10.3^{10.7}_{10.0}$ | $10.4^{10.7}_{10.0}$ |
| | $\bar{t}_{M-w}$ [Gyr] | $0.3^{0.7}_{0.1}$ | $0.7^{1.6}_{0.3}$ | $0.2^{2.3}_{0.1}$ | $0.1^{0.6}_{0.0}$ | $0.7^{1.7}_{0.4}$ |
| | $z_{form}$ | $2.7^{3.9}_{1.7}$ | $3.8^{6.0}_{2.3}$ | $3.6^{5.1}_{2.4}$ | $2.6^{3.8}_{1.9}$ | $3.9^{5.6}_{2.4}$ |
| | $z_{M,*=10^8 M_\odot}$ | $3.1^{4.8}_{1.8}$ | $6.6^{8.9}_{1.8}$ | $9.0^{9.9}_{3.2}$ | $3.0^{4.5}_{1.8}$ | $6.0^{8.9}_{3.3}$ |
| | $z_{M,*=10^9 M_\odot}$ | $3.0^{4.8}_{1.7}$ | $5.9^{8.3}_{3.7}$ | $4.4^{9.4}_{2.6}$ | $3.0^{4.6}_{1.8}$ | $5.7^{8.3}_{1.4}$ |
| | SFR [$M_\odot yr^{-1}$] | $30^{299}_{7}$ | $102^{594}_{12}$ | $32^{357}_{5}$ | $99^{1593}_{5}$ | $254^{1888}_{24}$ |

## 4. Results

### 4.1. Early mass build-up of massive galaxies

According to the fiducial model, the galaxies in our sample have a median stellar mass (and 16th and 84th percentiles) of $\log(M/M_\odot) = 10.5^{11.0}_{10.1}$, a median redshift of $z = 2.4^{3.5}_{1.6}$, a median mass-weighted age of $t̄_M = 0.7^{1.8}_{0.2}$ Gyr, and a formation redshift of $z_{form} = 3.9^{6.6}_{2.7}$. Moreover, they have assembled $10^8 M_\odot$ by redshift $z_8 = 6.7^{9.0}_{3.9}$ and $10^9 M_\odot$ at redshift $z_9 = 5.8^{8.3}_{3.8}$.

Figure 8 shows the formation, quenching, and observed times of the galaxies sample as a function of their stellar mass, according to our fiducial model. We observe a clear downsizing trend, wherein more massive galaxies tend to form and quench earlier in cosmic history. At fixed stellar mass, formation times shift toward higher redshift as stellar mass increases. The only galaxy with mass above $10^{11} M_\odot$ that is quenched does so by $z \sim 4$. This is consistent with the well-established picture in which massive galaxies undergo rapid early star formation and subsequently quench earlier than their lower-mass counterparts (Carnall et al. 2018b; Forrest et al. 2020a,b). Our results also align with the scenario described by Rinaldi et al. (2025a), wherein massive galaxies rapidly establish a self-regulated, steady state of star formation along the main sequence.

In Figure 8 we also plot the maximum observed stellar mass that would be expected for a given star formation efficiency $\epsilon$. We derive the maximum stellar mass as a function of the redshift for a given $\epsilon$ in the NIRISS footprint using the hmf python package (Murray et al. 2013) assuming an SMT fitting function (Sheth et al. 2001). The more reasonable upper limit on the observed stellar mass would use $\epsilon = 0.2$, the maximum inferred from the peak of the stellar mass to halo mass relation across a range of redshifts (Akins et al. 2024). At fixed stellar mass, galaxies that formed at higher redshifts are more compatible with a higher star formation efficiency. This is in agreement with the emerging picture that the assembly of halo and stellar mass has proceeded at different rates throughout cosmic history (e.g. Shuntov et al. 2025).

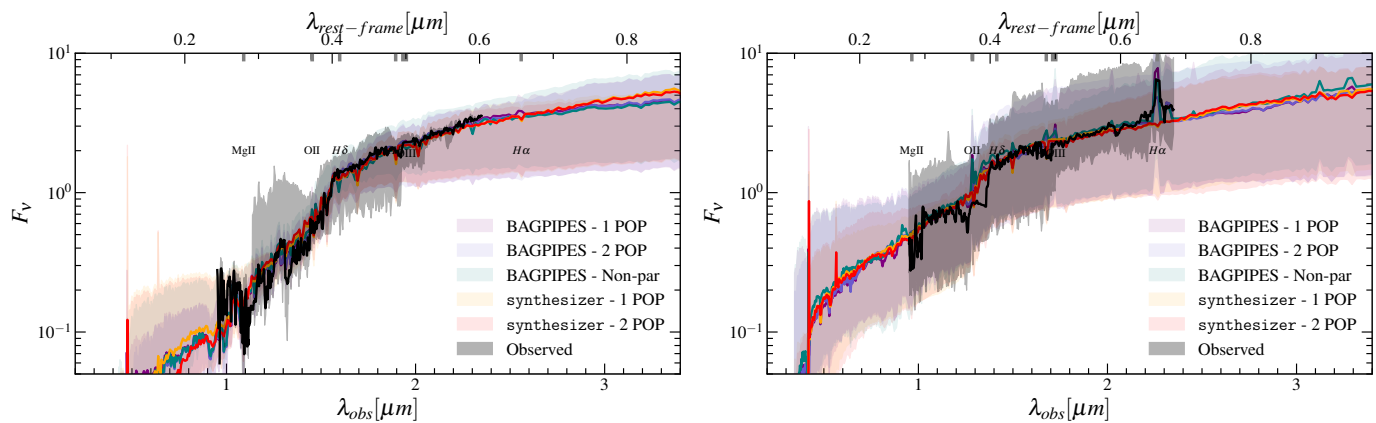

**Fig. 6.** Left panel: Median SEDs of quiescent galaxies obtained from the different runs, with the shaded regions indicating the 84th–16th percentile range. Overplotted are the stacked low-resolution spectra from NIRISS. Right panel: Same as the left panel, but for star-forming galaxies. All SEDs and spectra are redshifted to the median redshift of the corresponding sample.

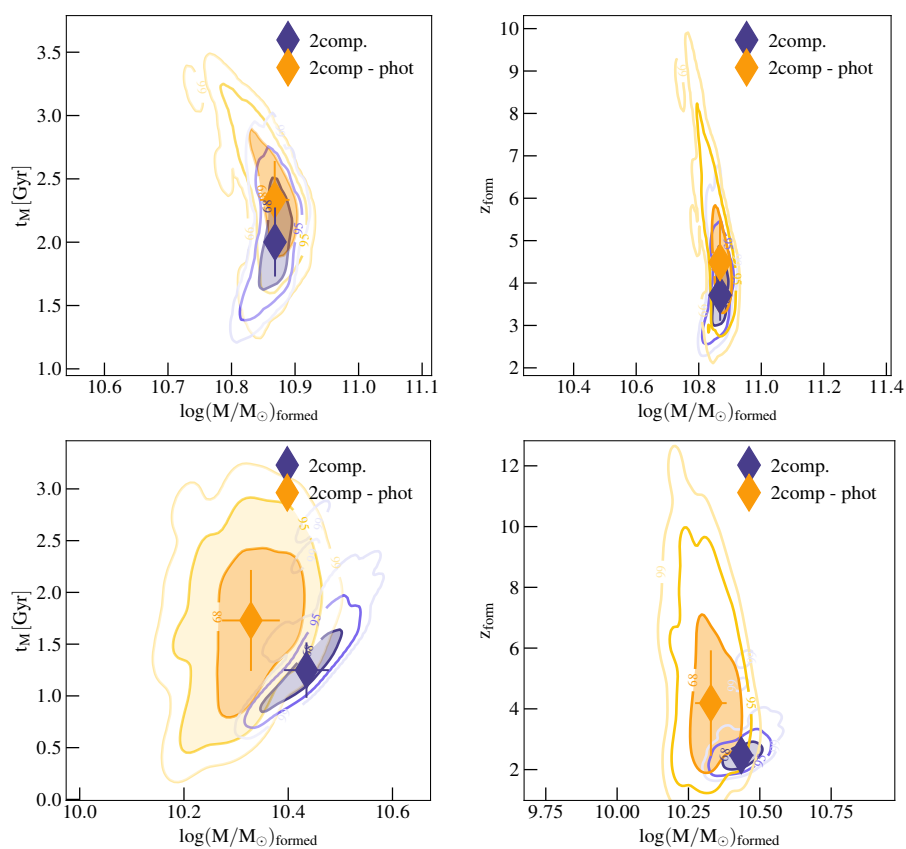

**Fig. 7.** Left panels: Distribution of $t_M$ versus stellar mass derived from the photometry-only run and the fiducial run for the two galaxies (top and bottom) shown in Figure 5. Right panels: Same as the left panels, but displaying the distribution of $z_{form}$ versus stellar mass.

Figure 9 presents the distribution of star formation histories (SFHs) for the full galaxy sample for the three BAGPIPES runs. Both the non-parametric run and the fiducial model result in very similar values of the median formation redshift (see Table 2). The non-parametric SFH exhibit an earlier onset of stellar mass assembly, with significant star formation already in place by $z \sim 9.6$, corresponding to a cosmic time of $\sim 0.52$ Gyr. In contrast, the median SFRs in both the 1-population and 2-population parametric models remain low until $z \sim 7.1$, or $\sim 0.75$ Gyr after the Big Bang, implying a delay of approximately 230 Myr in the onset of star formation relative to the non-parametric case.

We derive the redshifts at which each galaxy reaches a stellar mass of $10^8$ and $10^9 \, M_\odot$, as these quantities are independent of both the redshift of observation and the observed stellar mass. This approach enables a direct comparison with high-redshift studies of the galaxy stellar mass function (SMF; e.g., Weibel et al. 2024; Shuntov et al. 2025; Harvey et al. 2025). Using our fiducial model, we compute the comoving number densities of galaxies that have assembled at least $10^8 \, M_\odot$ and $10^9 \, M_\odot$ in stellar mass within the redshift interval $8.5 \leq z \leq 9.5$. We find number densities of $(8.2 \pm 2.1) \times 10^{-4} \, \text{Mpc}^{-3}$ and $(5.8 \pm 2.1) \times 10^{-4} \, \text{Mpc}^{-3}$, respectively. These values are consis-tent within $1\sigma$ with the SMF estimates at $z \sim 9$ from Weibel et al. (2024), who report a number density of $\sim 10^{-4}$ Mpc$^{-3}$ at $M_* \sim 10^9 M_\odot$. Our results also agree with the findings of Harvey et al. (2025), who infer similarly high number densities for galaxies at $10^8 M_\odot$ at $z > 8.5$, suggesting that efficient stellar mass assembly was already underway in a substantial fraction of the population at these early epochs. Furthermore, our inferred densities are in agreement with the abundance of $10^8$ and $10^9 M_\odot$ galaxies predicted from clustering and luminosity function modeling in Shuntov et al. (2025). These elevated number densities at high redshift also support the scenario of high star formation efficiencies at $z \geq 9$, in agreement with results from recent numerical simulations (Ceverino et al. 2024).

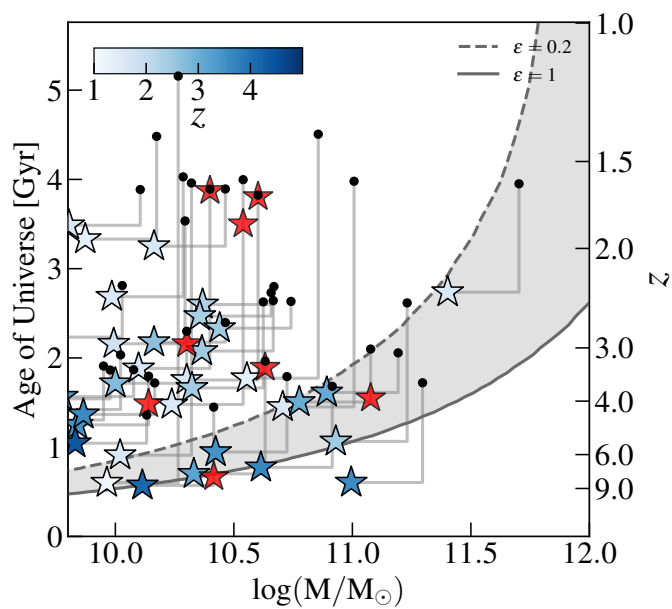

**Fig. 8.** Formation and quenching redshift for our sample as parametrized by BAGPIPES - 2POP. Blue-ish stars on the y-axis represent the age (and redshift) at which the galaxies form ∼ 50% of their stellar mass, with the corresponding 50% observed stellar mass on the x-axis. The stars are color-coded according to the redshift of the observation. Red stars indicate time of quenching and the observed stellar mass. Black circles denote the age of the Universe at the time of the observation for each galaxy (and the corresponding observed stellar mass). These points are connected by solid lines for each galaxy. We also overplot the maximum observed stellar mass that would be expected for a given star formation efficiency ($\epsilon$).

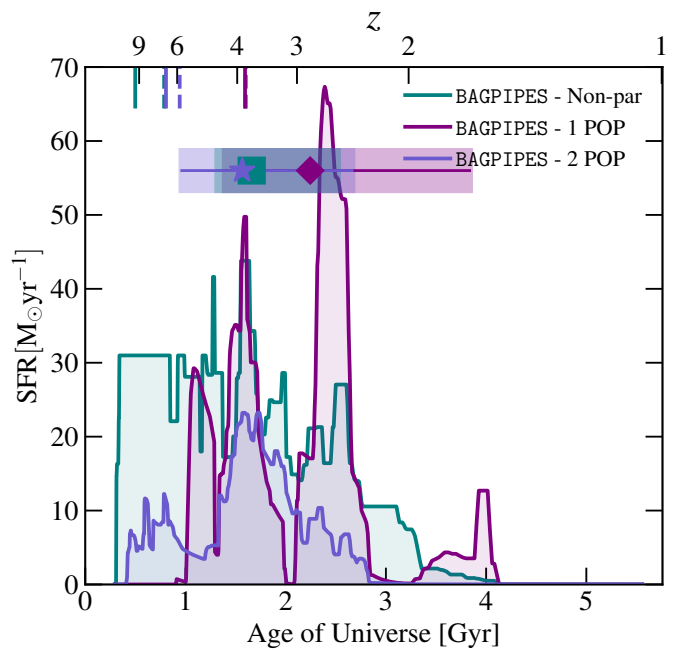

**Fig. 9.** Average SFHs of galaxies in our sample. The shaded colored areas correspond to the $16^{th}$- $84^{th}$ percentile interval computed from the scatter of the SFHs of all galaxies in our sample. The three points on top show the median formation redshifts, while the boxes extend to the $16^{th}$ and $84^{th}$ percentiles. Solid vertical lines show the median redshift at which this type of galaxies form $10^8 M_\odot$ according to the different models, while dashed vertical lines show the median redshift at which this type of galaxies form $10^9 M_\odot$.

### 4.2. Trends with redshift

Our galaxy sample spans a wide redshift range, which could potentially dilute some of the observed trends due to evolutionary effects. To mitigate this, we divide the sample into two redshift bins, $1 \leq z < 2$ and $z \geq 2$, which approximately correspond to similar intervals in cosmic time. Figure 10 shows the distribution of star formation histories (SFHs) for galaxies in each redshift bin, while the median physical properties of the two subsamples are summarized in Table 2.

On average, lower-redshift galaxies have a higher median mass-weighted age of $\bar{t}_M = 2.0^{1.2}_{3.0}$ Gyr and a formation redshift of $z_{form} = 3.4^{5.9}_{2.0}$. At these lower redshifts, the difference between non-parametric and parametric SFHs becomes particularly pronounced. Non-parametric SFHs predict that the galaxies reach $10^8$ and $10^9 M_{sun}$ at higher redshifts. The difference between $z_8$ and $z_9$ is also much higher (see Table 2 for non-parametric models, implying a slower mass assembly.

At higher redshifts, galaxies have a median mass-weighted age of $\bar{t}_M = 0.5^{0.9}_{0.1}$ Gyr and a formation redshift of $z_{form} = 4.0^{6.3}_{2.9}$. Interestingly, the differences between non-parametric and parametric star formation histories (SFHs) are less pronounced in this redshift regime. The offset in the onset of star formation is smaller, with $z_8$ 1.2× larger than the fiducial model . A detailed comparison of all codes and SFH assumptions is presented in Appendix A. At $\bar{z} = 0.8$, Kaushal et al. (2024) found that non-parametric models derived with `Prospector` (Leja et al. 2019) generally exhibit more extended SFHs than parametric ones generated with `BAGPIPES`. A similar trend was reported in Leja et al. (2019) at $\bar{z} = 0.07$, where non-parametric models consistently produced older stellar ages compared to parametric models from `BAGPIPES` and other fitting codes. This discrepancy is likely due to differences in how star formation onset is treated: parametric models typically allow the onset time to vary freely, whereas non-parametric models usually assume that star formation begins at $t = 0$. While previous works have identified similar trends, this is the first time that a consistent comparison is performed using the same fitting code but with different SFH assumptions. Furthermore, we present here the first evidence of a redshift-dependent trend in the divergence between parametric and non-parametric SFHs.

### 4.3. Trends with star formation activity

We divide our sample into star-forming galaxies (SF) and quiescent galaxies (Q) on the basis of the star-formation history of our fiducial model. All galaxies in the quiescent sample are also quiescent according to the UVJ diagram described in Section 3. Moreover, they have very low star formation rate (SFR; averaged over the last 100Myr) according to all the models. Quiescent galaxies have stellar mass of $\log(M/M_\odot) = 10.5^{10.6}_{10.3}$, mass-weighted age of $\tilde{t}_{M-w} = 1.1^{2.0}_{0.5}$ Gyr and formed at $z_{form} = 4.1^{7.4}_{3.5}$. Table 2 shows that quiescent galaxies are older with respect to the total and SF sample regardless of the choice of SFH or the code used.

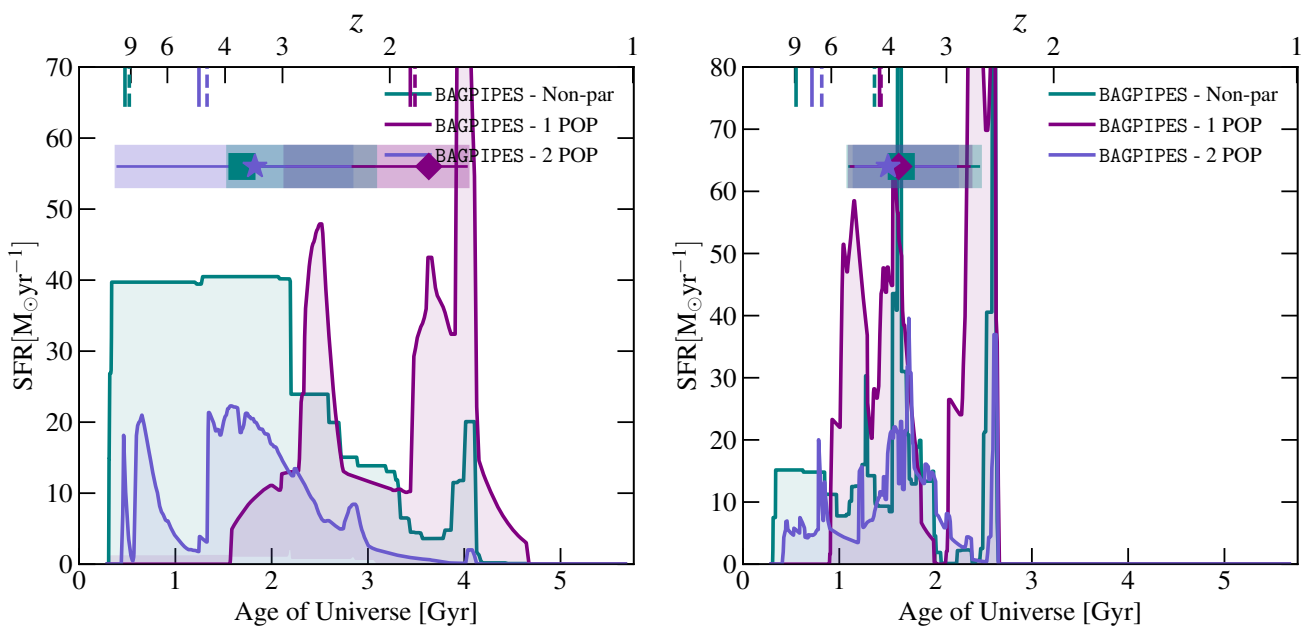

**Fig. 10.** Left panel: Average SFHs in the lowest redshift ($z < 2$). The shaded colored areas, points and lines have the same meaning as in Figure 9. Right panel: same as left panel but for galaxies at higher redshift ($z \geq 2$).

Figure 11 shows the average SFH of quiescent galaxies in our sample. According to our fiducial SED model, BAGPIPES - 2 POP, these galaxies assemble the first $10^8 M_\odot$ and $10^9 M_\odot$ very early in the Universe, with $z_8 = 6.8^{8.8}_{4.6}$ and $z_9 = 5.7^{8.0}_{4.4}$, respectively. They also assemble very rapidly, as shown by the small difference between $z_8$ and $z_9$. Our galaxies exhibit different quenching redshifts, i.e the redshift at which the star formation rate (SFR) falls below 10% of the average SFR throughout the galaxy's history Carnall et al. (2018b) as shown in Figure 8.

Unfortunately, we do not have enough quiescent galaxies in our sample to derive their mean properties in different redshift bins. However, in Figure 12 we plot the formation redshift of our quiescent galaxies as a function of stellar mass and with the points color-coded according to redshift. In the plot, we also show some of the most recent results on the assembly histories of massive quiescent galaxies (e.g. Belli et al. 2019; Carnall et al. 2018b; Beverage et al. 2025; Park et al. 2024; Nanayakkara et al. 2025; Slob et al. 2024). The higher redshift galaxies in our sample have a formation redshift of $5 \lesssim z_{form} \lesssim 9$ and a quenching redshift of $z_{quench} \sim 4$. The early onset of star formation observed in high-redshift quiescent galaxies indicates that the assembly of massive galaxies in the early Universe occurred at a significantly faster pace compared to their counterparts in the local Universe. Moreover, observations of quiescent galaxies at redshifts up to $z \sim 4-5$ (e.g., Forrest et al. 2020a; Carnall et al. 2023; Setton et al. 2024) indicate that the evolution of the star formation timescale extends to significantly earlier epochs.

We do not see any trend between formation redshift and stellar mass in our sample, although with a very limited number of galaxies.

SF galaxies constitute the dominant population within the studied sample. Figure 13 shows their average SFHs for to different runs. According to our fiducial model, SF galaxies have stellar masses of $\log(M/M_\odot) = 10.5^{11.0}_{10.1}$, mass-weighted ages of $\tilde{t}_{M-w} = 0.7^{2.0}_{0.3}$ Gyr. This makes them significantly younger on average compared to quiescent galaxies and form at a slightly later epochs, with $z_{form} = 3.8^{6.0}_{2.4}$.

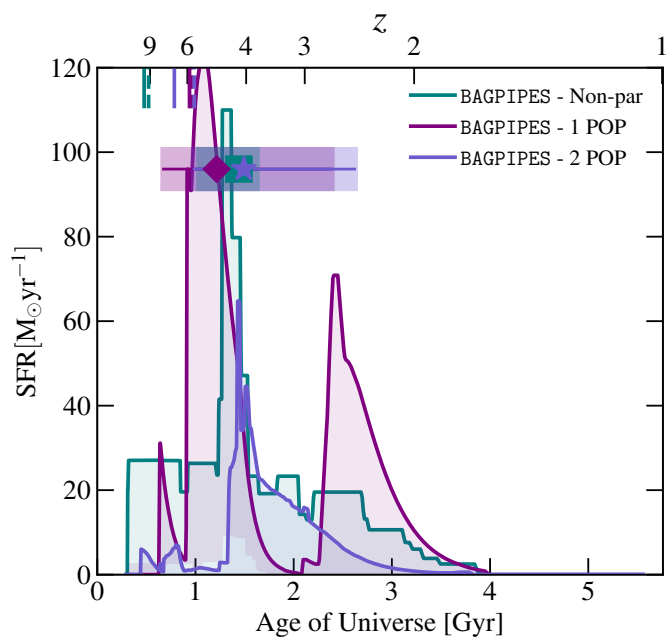

**Fig. 11.** SFHs of quenched galaxies according to different codes and SFHs. The shaded colored areas, points and lines have the same meaning as in Figure 9.

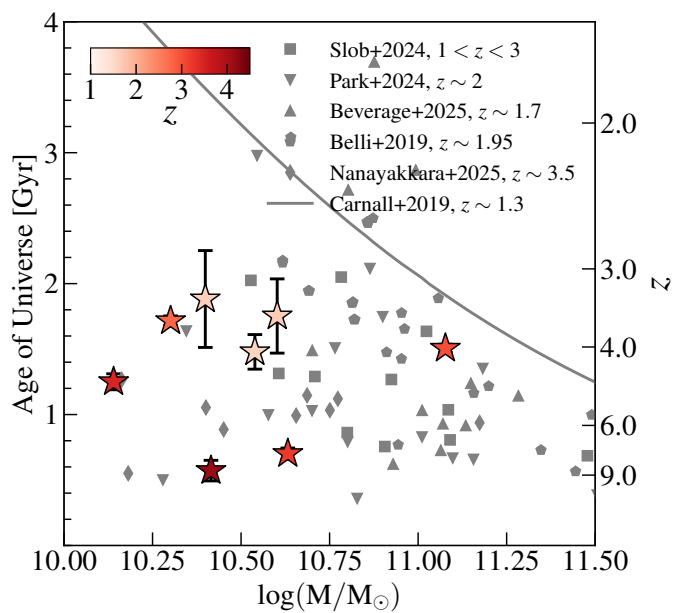

**Fig. 12.** The age of the Universe (on the left) and the redshift (on the right) at the time of formation for the quiescent galaxies in the sample are presented as a function of their stellar mass. he symbols are colored by the redshift at observation. We also add several recent results from literature at different redshifts as defined in the legend.

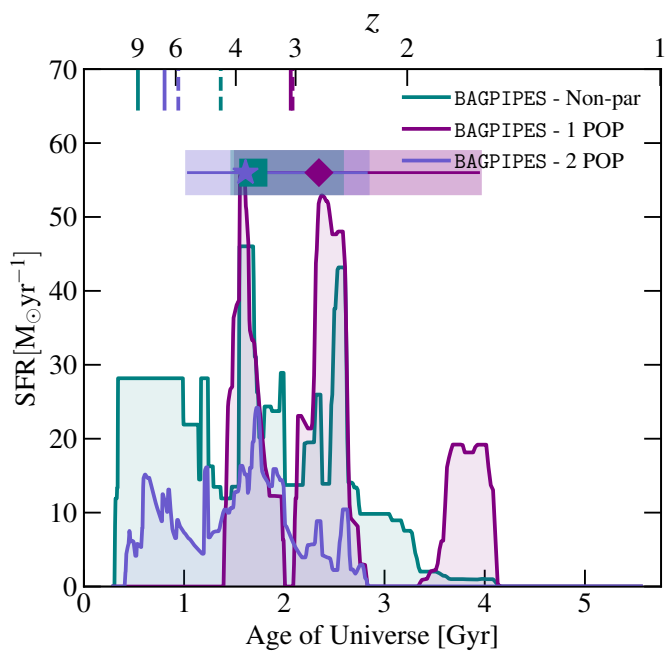

**Fig. 13.** SFHs of star-forming galaxies according to different codes and SFHs. The shaded colored areas, points and lines have the same meaning as in Figure 9.

## 5. Conclusions

In this paper, we present a comprehensive analysis of the star formation histories (SFHs) of a mass-complete sample of galaxies with stellar masses $\geq 10^{10} M_\odot$, spanning the redshift range $1 < z < 4.5$. Our analysis leverages a rich set of spectro-photometric data, including broadband photometry and low-resolution grism spectroscopy from JWST/NIRISS, enabling us to probe the stellar populations of galaxies across cosmic time with im-proved accuracy. By performing SED fitting with two modeling codes —`BAGPIPES` and `synthesizer`—and employing multiple SFH assumptions (parametric, two-component, and non-parametric forms), we critically assess how different methodological choices impact the derived formation timescales and assembly histories of massive galaxies. The complete comparison of these modeling approaches is presented in the main text and Appendix A, while in the following, we adopt as our fiducial model the results from `BAGPIPES` using a two-population SFH (see Section 3.2). Our main results include:

- Given the wide redshift range of our sample, we observe substantial diversity in stellar population ages. On average, galaxies in our sample exhibit mass-weighted stellar ages of $t_{M-w} \sim 0.8$ Gyr and median formation redshifts of $z_{form} \sim 3.9$. These values reflect the fact that even at relatively early cosmic times, massive galaxies had already assembled a substantial portion of their stellar mass.

- Galaxies assemble rapidly, they have formed $10^8 M_\odot$ and $10^9 M_\odot$ by $z_8 = 5.9$ and $z_9 = 5.4$, respectively.

- We divide the sample into two redshift bins: $1 \leq z < 2$ and $2 \leq z \leq 4.5$, and find notable differences in the stellar population properties between the two subsets. Despite having similar stellar masses, the two redshift subsamples trace different evolutionary stages. Galaxies observed at $1 \lesssim z < 2$ have mass-weighted ages of $t_{M-w} \simeq 2.0$ Gyr and formed half of their present-day stellar mass by $z_{form} \simeq 3.4$; this means that several gigayears have therefore elapsed between their main build-up and the epoch of observation. In contrast, galaxies observed earlier, at $2 \lesssim z \lesssim 4.5$, are only $t_{M-w} \simeq 0.5$ Gyr old, even though their formation redshifts are slightly higher ($z_{form} \approx 4.0$). The younger ages simply reflect the much shorter interval (less than a gigayear) between formation and observation. Thus, the apparent age gap arises naturally from look-back time: the farther back we look, the closer each galaxy lies to its own formation epoch.

- Our analysis highlights that non-parametric SFHs tend to favor earlier formation epochs compared to parametric ones. This is particularly pronounced in the lower-redshift subsample, where the non-parametric model implemented in `BAGPIPES` predicts that galaxies reach $10^8$ and $10^9$, $M_\odot$ by $z_8 = 9.8$ and $z_9 = 9.2$, respectively—much earlier than in parametric models. These findings emphasizes that the choice of SFH parameterization can significantly influence derived formation timescales.

- Although our sample includes a limited number of quiescent galaxies across the full redshift range, they exhibit systematically older stellar populations ($t_{M-w} \sim 1.1$ Gyr) and higher formation redshifts ($z_{form} \sim 4.1$) than the star-forming population. While the small number of quiescent systems precludes a detailed statistical analysis, we do observe a tentative trend whereby quiescent galaxies at higher redshift formed earlier, with formation redshifts reaching as high as $z_{form} \sim 9$. These results are qualitatively consistent with the "downsizing" scenario, in which the most massive and passive systems form earlier and more rapidly.

- In contrast, star-forming galaxies in our sample are typically younger ($t_{M-w} \sim 0.7$ Gyr) and formed more recently, with median formation redshifts of $z_{form} \sim 3.8$.

*Acknowledgements.* MA acknowledges financial support from Comunidad de Madrid under Atracción de Talento grant 2020-T2/TIC-19971. This work has made use of the Rainbow Cosmological Surveys Database, which is operated by the Centro de Astrobiología (CAB/INTA), partnered with the University of California Observatories at Santa Cruz (UCO/Lick.UCSC). The project that gave rise to these results received the support of a fellowship from the "la Caixa" Foundation (ID 100010434). The fellowship code is LCF/BQ/PR24/12050015. LC acknowledges support from grants PID2022-139567NB-I00 and PIB2021-127718NB-I00 funded by the Spanish Ministry of Science and Innovation/State Agency of Research MCIN/AEI/10.13039/501100011033 and by "ERDF A way of making Europe".

## Appendix A: Dependence of Results on A Priori Assumptions and Fitting Techniques

In this section, we describe the robustness of the different parameters that can be obtained from the SED fitting analysis. To understand better the differences shown in Table 2, we compare the parameters obtained in the different runs against the value of the "fiducial" model. Throughout this work, we assume the `BAGPIPES` - 2 POP run as our fiducial model.

### A.1. Stellar masses

The mass differences of $log(M_*/M_{*,fiducial})$, with respect to the fiducial mass, are shown in Figure A.1. Over the entire mass range, the median differences between the mass obtained with different codes and the fiducial mass is up tp ~0.1 dex for all runs. This is consistent with previous comparisons between different codes (e.g. Leja et al. 2019; Pacifici et al. 2023). As expected, stellar mass is a well constrained parameter that can be recovered by different codes and different assumptions. There are not any observable trends, either with stellar masses or with different codes.

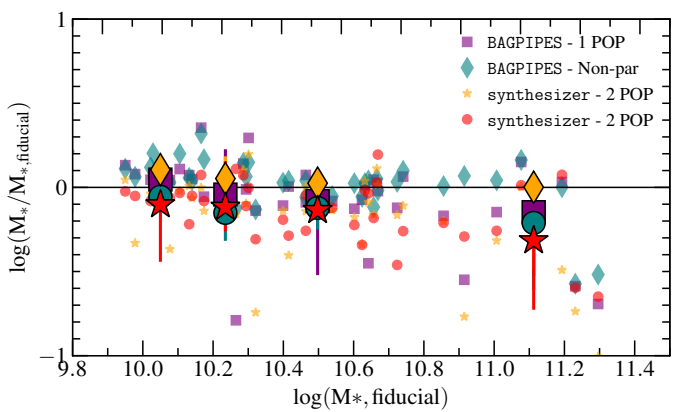

**Fig. A.1.** Stellar mass differences (in log) for the different SFH models and the different codes with respect to the fiducial one. Smaller points are individual galaxies, bigger points with the error bars are the median obtained in each bin. The bins have been chosen to contain the same number of galaxies. Shaded regions represent $1\sigma$ errors.

### A.2. Mass-Weighted ages and formation redshifts

Figure A.2, shows the comparison between the mass-weighted ages and formation redshifts of the different codes and assumptions against the fiducial model. Over the entire sample, both `BAGPIPES` and `synthesizer` with a delayed$-\tau$ model, predicts smaller mass-weighted ages with respect to the fiducial model. This difference is of the order of ~ 47% for `BAGPIPES` - 1 POP and ~65% for `synthesizer` - 1 POP. This is consistent with the fact that single population models might be biased towards the youngest stars which can overshine the more evolved stellar population. `Synthesizer` - 2 POP predicts a difference in mass weighted ages of 12% with respect to the fiducial model, the lowest difference between our runs. As our fiducial model is the BAGPIPES with two populations, this suggest that the assumptions on the SFH are more important than the intrinsic differences from different codes. `BAGPIPES` - Non-par also agrees very well with the fiducial model, with an average difference of 26%.

Formation redshifts take into account the different redshifts at which our galaxies are observed. From Figure A.2, we observe similar trends as for the mass-weighted ages. However, this parameter is better constrained. `BAGPIPES` - 1 POP gives $\Delta z_{form}/(1 + z_{form})$ of ~ 11%, while `synthesizer` - 1 POP of ~ 13%. `Synthesizer` - 2 POP has the better agreement with the fiducial model, with $\Delta z_{form}/(1 + z_{form}$ of ~ of 3% over the whole sample. A good agreement ($\Delta z_{form}/(1 + z_{form} \sim$ -5%) is found also for `BAGPIPES` - Non-par.

### A.3. Formation timescales

In order to compare how rapidly our galaxies form stars, we derive the times at which the galaxies have formed their first $10^8 M_\odot$ and $10^9 M_\odot$. Figure A.3 shows the comparison between $t(10^8 M_\odot)$ and $t(10^9 M_\odot)$ versus the fiducial model. Also in this case, the best agreement is between `synthesizer` - 2 POP and the fiducial model with a median difference of 9% and 1% for $t_8$ and $t_9$. A 15% difference in $t_8$ is found for `BAGPIPES` - Non-par, but with a much larger scatter. The two single population models from `BAGPIPES` and `synthesizer` consistently predicts lower $t_8$ and $t_9$, with a difference of ~ 60% for `BAGPIPES` and ~ 70% `synthesizer`. Figure A.4 shows the same results as in Figure A.3, but in terms of redshift. While the formation redshift of the galaxy is mostly independent on the choice of star-formation history and for different codes, A.4 shows that when a non-parametric SFH is considered, galaxies start to assemble their stellar mass at earlier epochs. The discrepancy between Non-parametric SFHs and the parametric models is particularly evident at low $z_8$ and $z_9$. On average, `BAGPIPES`- Non-parametric has $\Delta z_8/(1 + z_8)$ of ~ 40% and $\Delta z_9/(1 + z_9)$ of 34% with respect to the fiducial model.

### A.4. Star formation rate

We derived the star-formation rate averaged over the last 100 Myrs with the different prescriptions for the SED fitting. Figure A.5, shows the comparison between all models and the SFR derived from UV and mid-IR data (Barro et al. 2019), in this case the 'true' SFR. `BAGPIPES` - 1 POP and `BAGPIPES` - 2 POP underestimate the 'true' SFR by a factor of 40% . `BAGPIPES` - Non-par underestimates the 'true' SFR by a factor of 13%. `Synthesizer` - 1 POP underestimates the 'true' SFR by a factor of 10%. `Synthesizer` - 2 POP is the only model that overestimates the 'true' SFR by a factor of 49%.

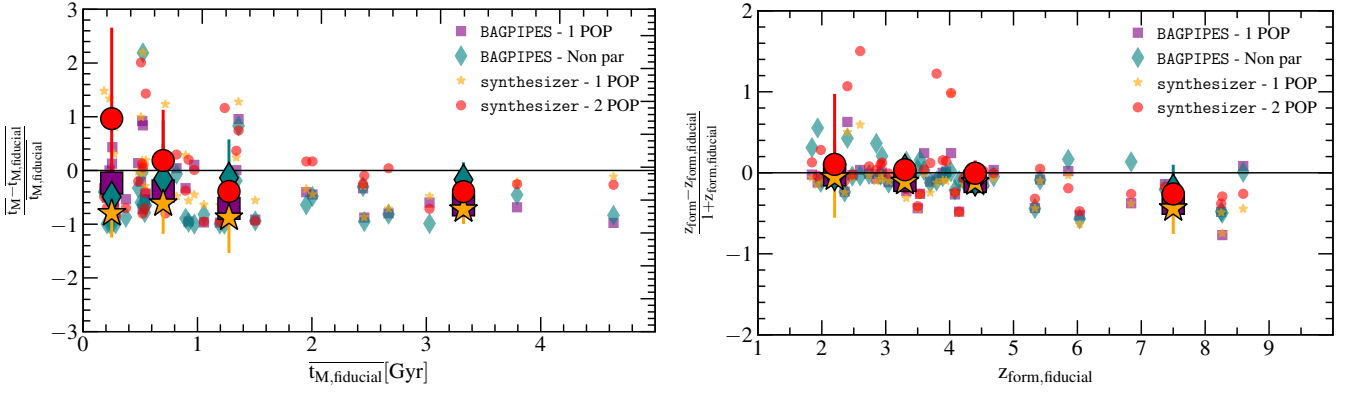

**Fig. A.2.** Left panel: Relative difference between the mass-weighted ages derived by the different codes with respect to the fiducial one. Right panel: same as left panel but for the formation redshift. In both panels, smaller points are individual galaxies, bigger points with the error bars are the median obtained in each bin. The bins have been chosen to contain the same number of galaxies.

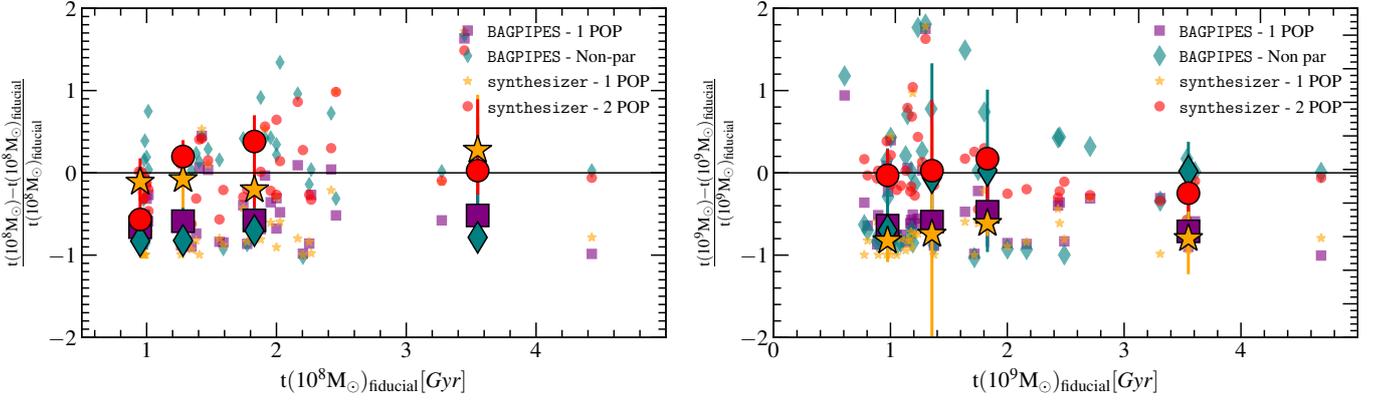

**Fig. A.3.** Left panel: Relative difference between the time at which the galaxy assembles its first $10^8 M_\odot$ with respect to the fiducial one (synthesizer - 2 POP). Right panel: same as left panel but for $t(10^9 M_\odot)$. In both panels, smaller points are individual galaxies, bigger points with the error bars are the median obtained in each bin. The bins have been chosen to contain the same number of galaxies.

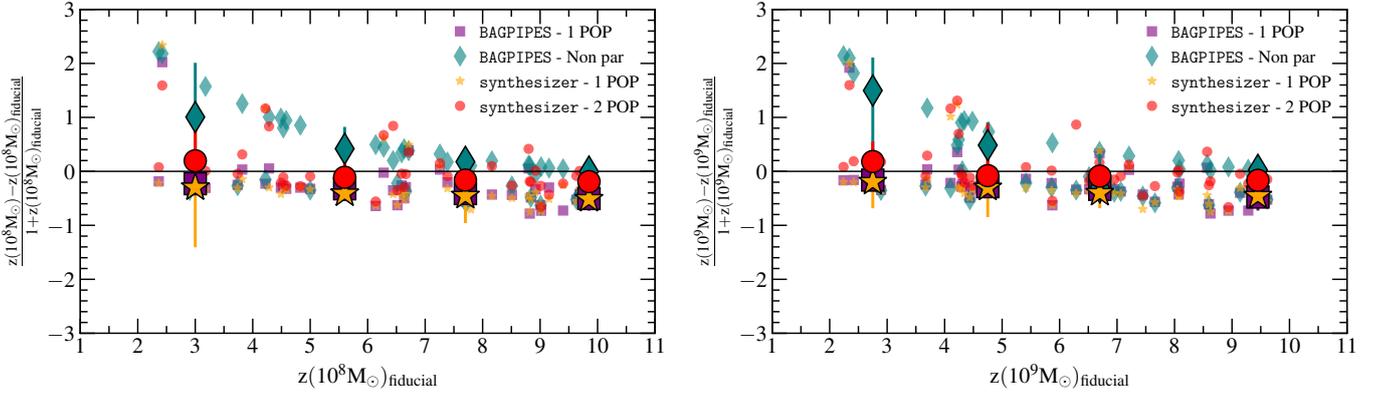

**Fig. A.4.** Left panel: Relative difference between the redshift at which the galaxy assemble $10^8 M_\odot$ with respect to the fiducial one (synthesizer - composite). Right panel: same as left panel but for $z_9$. In both panels, smaller points are individual galaxies, bigger points with the error bars are the median obtained in each bin. The bins have been chosen to contain the same number of galaxies.

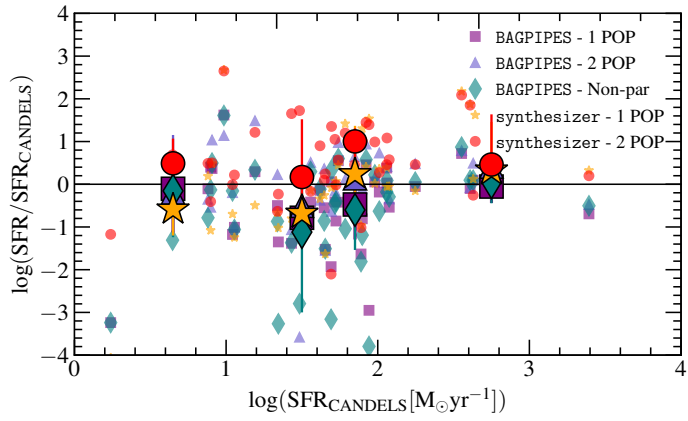

**Fig. A.5.** SFR differences (in log) for the different SFH models and that estimated from the combination of UV and mid-IR data. Smaller points are individual galaxies, bigger points with the error bars are the median obtained in each bin. The bins have been chosen to contain the same number of galaxies.